\documentclass[12pt]{iopart}
\usepackage{graphicx}% Include figure files
\graphicspath{{figure/}}
\usepackage{color}
\usepackage{bm}
\usepackage{dcolumn}% Include figure files
\usepackage{latexsym} 

\newcommand{\mbfx}{{\bf x}}
\newcommand{\mbfy}{{\bf y}}

\newcommand{\D}{\,\mathrm{d} }

\begin{document}

%\draft
%\title{Fluctuations and non-Gaussianity in 3D+1 quantum nonlocal nonlinear waves: towards quantum gravity and quantum fluid technologies}
\title{Random walk and non-Gaussianity of the 3D second-quantized Schr\"odinger-Newton nonlocal soliton}
\author{Claudio Conti}
\address{Department of Physics, University Sapienza, Piazzale Aldo Moro 5, 00185 Rome, Italy}
\address{Institute for Complex Systems, National Research Council (ISC-CNR), Via dei Taurini 19, 00185 Rome, Italy}
\address{Research Center Enrico Fermi, Via Panisperna 89a, 00184 Rome, Italy}
\ead{claudio.conti@uniroma1.it}
\begin{indented}
  \item[]{\today}
  \end{indented}

\begin{abstract}
  Nonlocal quantum fluids emerge as dark-matter models and tools for quantum simulations and technologies.
  However, strongly nonlinear regimes, like those involving multi-dimensional self-localized solitary waves,
  are marginally explored for what concerns quantum features. We study the dynamics of 3D+1 solitons in the second-quantized nonlocal nonlinear Schr\"odinger-Newton equation. We theoretically investigate the quantum diffusion of the soliton center of mass and other parameters, varying the interaction length.
  3D+1 simulations of the Ito partial differential equations arising from the positive P-representation of the density matrix validate the theoretical analysis.
  The numerical results unveil the onset of non-Gaussian statistics of the soliton,
  which may signal quantum-gravitational effects and be a resource for quantum computing. The non-Gaussianity arises from the interplay between the soliton parameter quantum diffusion and the stable invariant propagation. The fluctuations and the non-Gaussianity are universal effects expected for any nonlocality and dimensionality.
\end{abstract}

\vspace{2pc}

\maketitle

\section{Introduction}
Three-dimensional (3D) self-localized nonlinear waves enter various fields of research~\cite{KivsharBook,malomed2021multidimensional}, but their quantum properties are unexplored. Classical three-dimensional solitary waves (in short, 3D solitons) need to be stabilized against catastrophic collapse. Nonlocality is a well-known mechanism for the stabilization~\cite{Turitsyn85, Garcia2000, Bang02} and nonlocal soliton are a fascinating research direction involving long-range Bose-Einstein condensates (BECs)~\cite{weitz2010, Carusotto2013, Calvanese14, defenu2021longrange}, boson stars~\cite{Dell2000} and dark-matter models~\cite{Paredes2016, garnier2021incoherent}. However, a mean-field description that overlooks quantum effects provides limited information on the dynamics of self-trapped multidimensional waves. This limitation is specifically relevant as recent investigations suggest the solitons as non-classical sources for quantum technologies and fundamental studies~\cite{Conti2014, Liang2018, Villari2018, Malomed20, conti2022a, RKL2022}. Results in 1D~\cite{Folli10, Batz2011} suggest that nonlocality frustrates fluctuations. However, despite ab-initio investigations on long-range interactions~\cite{d_drummond_quantum_2016, Deuar2017}, the quantum statistics of self-trapped 3D nonlocal solitons is an open issue.

In addition, recent work on gravitational interaction in BEC predicts non-Gaussian statistics~\cite{Howl2021}. Non-Gaussianity is a resource for continuous-variable quantum information science~\cite{Paris2014, Shapiro18} and its understanding in quantum fluids may enable new universal quantum processors. Also, emerging of non-Gaussian statistics in table-top experiments may open the way to study - or simulate - quantum gravity in the laboratory. Ref.~\cite{Howl2021} predicts that a BEC in a trap, once prepared in a squeezed state or Schr\"odinger-cat state, triggers the non-Gaussian statistics measured by a signal-to-noise ratio (SNR) parameter, which reveals quantized gravity. However - so far - no experiments or numerical simulations validate these theoretical predictions. Also, quantum fluctuations and non-Gaussianity in multidimensional self-trapped solitonic nonlocal condensates have never been considered before.

Here, we study theoretically and numerically the quantum dynamics 3D nonlocal solitons. We use a perturbative approach and we analytically predict the quantum diffusion of the soliton position and other parameters. We validate our analytical results by ab-initio numerical simulations based on the 3D+1 positive P-representation~\cite{Drummond2004, d_drummond_quantum_2016, Deuar2017}. We compute the SNR parameter introduced in~\cite{Howl2021}, which shows that non-Gaussianity arises in the quantum dynamics of 3D+1 nonlocal solitons, starting from a coherent state.
\section{Model and scaling}
We consider the many-body Hamiltonian
%  \begin{widetext}
\begin{equation}
\fl  \hat{H}=\frac{\hbar^2}{2m}\int \nabla\hat{\psi}^\dagger\cdot\nabla\hat{\psi} \D^3 \mbfx
  +\int U(\mbfx-\mbfx')\hat{\psi}^\dagger(\mbfx')\hat{\psi}(\mbfx)^\dagger\hat{\psi}(\mbfx')\hat{\psi}(\mbfx)\D^3 \mbfx \D^3\mbfx' \;,
  \label{eq:Poisson}
  \end{equation}
%\end{widetext}
with $m$ is the boson mass, and $U$ is the interaction potential.
We adopt the phase-space representation methods~\cite{GardinerBook,d_drummond_quantum_2016} for studying the nonlocal interaction.  The quantum field model is equivalent to a Fokker-Planck equation, which is mapped to Ito nonlinear partial differential equations coupling two fields $\psi$, and $\psi^+$
\begin{equation}
  \label{eq:nonlocalSDE}
  \begin{array}{lll}
\imath \hbar\partial_t\psi\,&=&-\frac{\hbar^2}{2m} \nabla^2\psi +\psi\,U*\rho+\sqrt{\imath \hbar }\psi\,\xi_U\\[4pt]
-\imath \hbar\partial_t\psi^+&=&-\frac{\hbar^2}{2m} \nabla^2\psi^+ +\psi^+\,U*\rho+\sqrt{-\imath\hbar }\psi^+\,\xi_U^+
  \end{array}
\end{equation}
where the asterisk denotes a convolution integral. In~(\ref{eq:nonlocalSDE}) $\rho=\psi^+\psi$, $\xi_U$ and $\xi_U^+$ are independent noises such that
\begin{equation}
  \label{}
  \langle \xi_U(\mbfx, t) \xi_U(\mbfx', t')\rangle =\langle \xi_U^+(\mbfx, t) \xi_U^+(\mbfx',t')\rangle =U(\mbfx-\mbfx')\delta(t-t')\;.
\end{equation}
The total number of particles is $\int \psi \psi^+ \D V$, its mean value is $N_T = \int \langle \psi \psi^+\rangle \D V$;
the brackets here denote the mean-field solution obtained with $\xi_{U}=\xi_{U}^{+}=0$.

In our numerical calculations below, we consider self-gravitating screened potential $U=-G m^2e^{-r/\Lambda}/r$, where $\Lambda$ is the interaction length. $G$ measures the coupling corresponding to the gravitational constant, but $U$ also models other long-range interactions as, e.g., thermal effects in photonic BEC~\cite{Calvanese14}. The mean-field theory is obtained by $\xi_U=\xi^+_U=0$, and $\psi^+=\psi^*$, and corresponds to the Schr\"odinger-Newton equation~\cite{RUFFINI1969}.
 \begin{figure*}[hbt]
  \includegraphics[width=\textwidth]{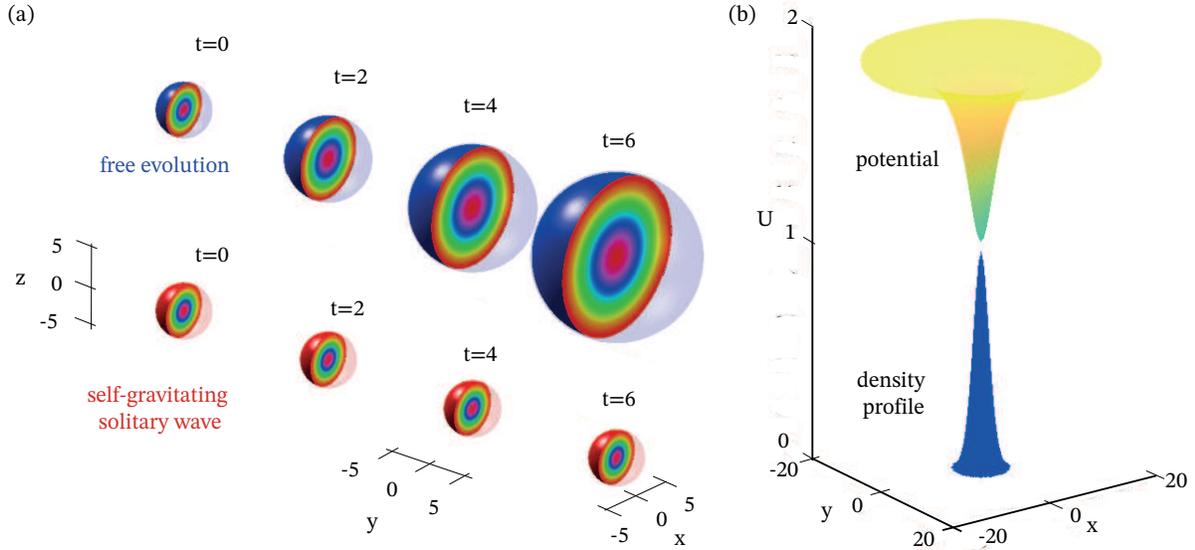}
  \caption{Self-gravitating solitonic core. (a) Comparison of time dynamics with and without interaction ($G=0$); 3D isodensity surfaces at different instants for freely evolving fields (top panel) and in the presence of the nonlinearity (bottom panel) with the time-invariant self-trapped wave-packet.  (b) Two-dimensional projection (average in the $z-$direction) of the density profile (blue) and resulting long-range potential (yellow)~\label{fig:solitonfeatures}}
\end{figure*}
\begin{figure*}[hbt]
  \centering
  \includegraphics[width=\textwidth]{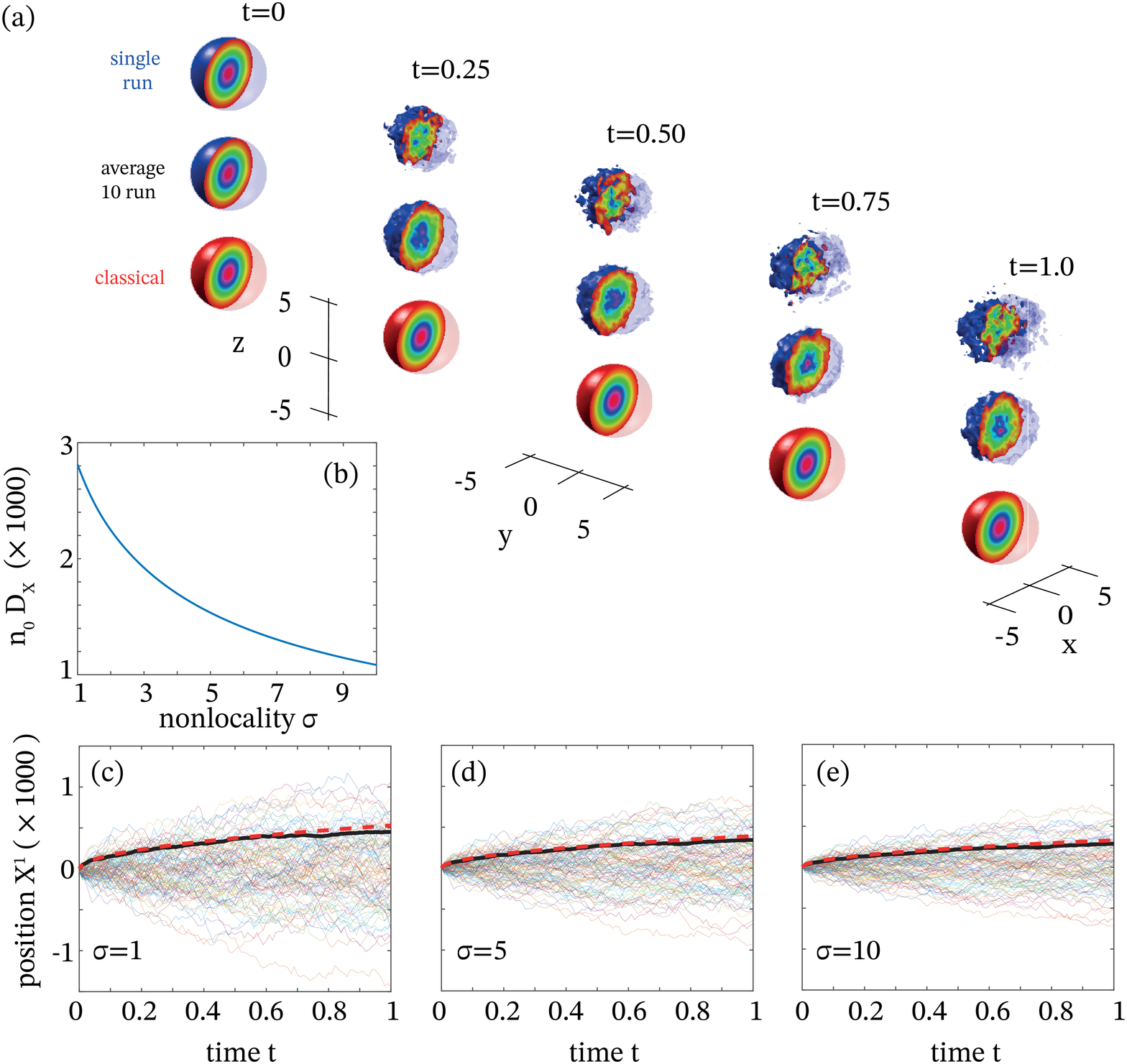}
  \caption{Classical and quantum evolution of the 3D+1 nonlocal soliton.  (a) Isosurfaces of the density $\rho$ of the solitonic core at different instants. We show a single run (top panel), an average of $10$ runs, and the classical propagation invariant solution (bottom panel).   (b) Diffusion coefficient $D_X$ after (\ref{eq:X2})  for various degrees of nonlocality $\sigma$.  (c,d,e) Computed trajectories for the $X^1(t)$ displacement for $100$ runs for three values of $\sigma$. The thick line is the standard deviation $\langle X^1(t)^2\rangle^{1/2}$, the dashed line is  (\ref{eq:X2}) for comparison with theory without fitting parameters.~\label{fig:solitonfluctuactions}}
\end{figure*}

We write the stochastic equations in dimensionless units by letting
\begin{eqnarray*}
  (x,y,z)&\rightarrow& (x,y,z) r_0\\
  t&\rightarrow& t t_0\\
\left(\psi,\psi^+\right)&\rightarrow& \psi_0 \left(\psi,\psi^+\right)
\end{eqnarray*}
and
\begin{eqnarray}
  t_0&=&2m r_0^2/\hbar\;, \\
  \psi_0^2&=&\frac{\hbar^2}{2 G m^3 r_0^4}\;,  \\
  n_0&=&\frac{\hbar^2}{2G m^3 r_0}=\frac{N_T}{M}
\label{eq:1}         
\end{eqnarray}
being
\begin{equation}
  M=\langle \int \psi\psi^+ \D^3\mbfx\rangle\;.
 \label{eq:2} 
\end{equation}
$n_{0}$ measures the number of particles in the condensate in units of $M$, the norm of the numerically obtained bound state profile [Eq.~(\ref{eq:boundstate}) below].

In the dimensionless units, Eqs.~(\ref{eq:nonlocalSDE}) read
\begin{equation}
  \label{eq:A:nonlocalSDE}
  \begin{array}{lll}
+\imath \partial_t\psi+\nabla^2\psi -\psi\,U\ast\rho&=s\\
 -\imath \partial_t\psi^++\nabla^2\psi^+ -\psi^+\,U\ast\rho&=s^+
  \end{array}
\end{equation}
with
\begin{equation}
U=-\frac{e^{-r/\sigma}}{r}\;,
\end{equation}
$\sigma=\Lambda/r_0$, and
\begin{equation}
  \begin{array}{lll}
    s=\sqrt{\frac{\imath}{n_0}}\xi_U\psi\;,\\[4pt]
    s^+=\sqrt{\frac{-\imath}{n_0}}\xi^+_U\psi^+\;.
    \end{array}
  \end{equation}
$\xi_U(x,y,z,t)$ and $\xi_U^+(x,y,z,t)$ are uncorrelated noise terms such that
  \begin{equation}
    \begin{array}{lll}
      \langle \xi_U(\mbfx,t)\xi_U(\mbfx',t')\rangle =\langle \xi^+(\mbfx,t)\xi^+(\mbfx',t')\rangle=
      U(\mbfx-\mbfx')\delta(t-t').
    \end{array}
  \end{equation}
% The stochastic equations in dimensionless units read as
% \begin{equation}
%   \label{eq:A:nonlocalSDE}
%   \begin{array}{lll}
% +\imath \partial_t\psi+\nabla^2\psi -\psi\,U\ast\rho&=&s\\
%  -\imath \partial_t\psi^++\nabla^2\psi^+ -\psi^+\,U\ast\rho&=&s^+
%   \end{array}
% \end{equation}
% with
% \begin{eqnarray}
%   U=-\frac{e^{-r/\sigma}}{r}\;,\\
%   s=\sqrt{\frac{\imath}{n_0}}\,\xi_U\psi\;,\\
%   s^+=\imath\sqrt{\frac{\imath}{n_{0}}}\,\xi^+_U\psi^+.\end{eqnarray}
$\sigma$ and $n_0$ are the dimensionless interaction length and particle number, respectively.

According to eq.~(\ref{eq:1}), one can either fix $r_{0}$ or $n_{0}$ to set all the other normalization constants.
  We choose to use $n_{0}$ because it appears explicitly in the normalized equation in a way such that the limit $n_{0}\rightarrow\infty$ corresponds to the mean-field regime. Indeed, the total mean particle number is $N_{T}=n_{0}M$.

  Once we have the numerical solution of the bound state [Eq.~(\ref{eq:boundstate}) below], which is determined by the scale $\sigma$, we study its quantum fluctuations by numerically solving Eqs.~(\ref{eq:A:nonlocalSDE}). In this paper, we fix a specific value for $n_{0}$, which allows us to perform numerical simulations with unitary time-scale in our normalized scale  (i.e., $t\simeq 1$ as in figure~\ref{fig:solitonfluctuactions} ), and we study the effects of a varying interaction length $\sigma$.

%%%%%%%%%%%%%%%%%%%%%%%%%%%%%%%%%%%%%%%%%%%%%%%%%%%%%%%%%%%%%%%
\section{Self-gravitating non-local soliton}
In the mean-field theory, equations~(\ref{eq:A:nonlocalSDE}) admit a stable radially-symmetric bound-state solution: a self-localized three-dimensional solitary wave. We write the solution with a Galileian boost as
\begin{equation}
  \label{eq:solitonansatz}
\psi=u(x^a-X^a)\exp\left[\imath \theta -\imath E t+\frac{i}{2}V^a(x_a-X_a)\right]\;,
\end{equation}
$\psi^+=\psi^*$, with $a=1,2,3$, $x^1=x$, $x^2=y$, $x^3=z$, omitting the sum symbol over repeated Latin indices. $u(x^a)$ is the real-valued soliton profile, such that
\begin{equation}
  \label{eq:boundstate}
\Delta u-U\ast u^2 u=Eu\;.
\end{equation}
The soliton energy $E$ is time-independent. For the position $X^a=X^a(t)$, we have (dot is the time-derivative)
\begin{equation}
  \label{eq:dynamic0}
  \begin{array}{lll}
    \dot X^a&=&V^a\\
    \dot V^a&=&0\\
    \dot \theta &=&\frac{1}{4}V^2,
    \end{array}
\end{equation}
with $V^2=\delta_{ab}V^aV^b$, and $\delta_{ab}$ the Kronecker symbol ($b=1,2,3$).
Equations~(\ref{eq:dynamic0}) imply
\begin{equation}
  \begin{array}{lll}
    X^a&=X^a(t)=X^a(0)+V^a t\;,\\
    \theta&=\theta(t)=\theta(0)+\frac{1}{4}V^2\;.
    \end{array}
  \end{equation}
Figure~\ref{fig:solitonfeatures} shows the evolution of the time-invariant soliton profile obtained from Eq.~(\ref{eq:boundstate}), compared with the evolution in the absence of nonlinearity ($U=0$). We also show the soliton compared with corresponding potential $U\ast \rho$.  The field profile and the potential are computed numerically. We use a pseudo-spectral parallel relaxation procedure in a 3D Cartesian domain. Figure~\ref{fig:solitonfeatures} shows the calculated classical bound state $u$. In the absence of interaction, the mass spreads upon evolution. In the presence of self-attraction, the solitonic wave packet is invariant upon propagation.
\section{Quantum effects on the 3D nonlocal soliton}
In the quantum regime, with $\xi_U\neq 0$ and $\xi_U^+\neq 0$, the soliton, initially prepared in a coherent state, evolves with fluctuations depending on the interaction length $\sigma$. The 3D+1 stochastic partial differential equations in~(\ref{eq:A:nonlocalSDE}) are solved by following Drummond and coworkers~\cite{d_drummond_quantum_2016}. We adopt an iterative stochastic solver with pseudospectral discretization and parallelized with the FFTW~\cite{FFTW05} and the Message Passing Interface (MPI) protocol. Figure~\ref{fig:solitonfluctuactions}a shows the numerical solution of the stochastic equations~(\ref{eq:A:nonlocalSDE}), which unveils that the soliton undergoes a random walk (figure~\ref{fig:solitonfluctuactions}c-e).

To study the quantum regime, we derive equations for the soliton parameters by~(\ref{eq:A:nonlocalSDE}) using soliton perturbation theory.
In the presence of noise, Eqs.~(\ref{eq:dynamic0}) are replaced by stochastic differential equations, which we derive by introducing the vectorial notation
\begin{equation}
{\bm \psi}=\left(\begin{array}{c}\psi\\\psi^+\end{array}\right)\;.
\end{equation}
Equations~(\ref{eq:A:nonlocalSDE}) are written as
\begin{equation}
\imath\sigma_3\partial_t {\bm \psi}+\Delta{\bm \psi}-U\ast(\psi\psi^+){\bm \psi}={\bm s}\;,
\end{equation}
with the Pauli matrix
\begin{equation}
\sigma_3=\left(\begin{array}{ll}1 & 0\\0 & -1\end{array}\right)\;,
\end{equation}
and
\begin{equation}
{\bf s}=\left(\begin{array}{c}s\\s^+\end{array}\right)\;.
\end{equation}
We introduce the following vector
\begin{equation}
  {\bf e}=\left(\begin{array}{lll}e^{\imath \theta -\imath E t+\frac{i}{2}V^a(x_a-X_a)}\\
  e^{-\imath \theta +\imath E t-\frac{i}{2}V^a(x_a-X_a)}\end{array}\right)\;,
\end{equation}
being $u=u(x^a-X^a,E)$ solution of Eq.~(\ref{eq:boundstate}). We also define
\begin{eqnarray}
  {\bf f}_\theta&=&{\bf e}u\\
{\bf f}_E&=&\imath\sigma_3{\bf e}\frac{\partial u}{\partial E}\;,\\
{\bf f}_{X}^a&=&-\imath\sigma_3{\bf e}\frac{\partial u}{\partial x^a} \;,\\
  {\bf f}_{V}^a&=&{\bf e}\frac{1}{2}(x^a-X^a)u \;.
                   \label{eq:vectors}
\end{eqnarray}
We introduce a scalar product for two vectors {\bf f} and {\bf g} such that
\begin{equation}
\left({\bf f},{\bf g}\right)=2\Re\int {\bf f}^*\cdot {\bf g} \D V\;.
\end{equation}
By using this scalar product, we build a bi-orthogonal system by
introducing the conjugate vectors to~(\ref{eq:vectors})
\begin{eqnarray}
  \hat{\bf f}_\theta&=&\imath \sigma_3{\bf f}_E\\
\hat{\bf f}_E&=&-\imath \sigma_3{\bf f}_\theta\\
\hat{\bf f}_{X}^a&=&\imath\sigma_3 {\bf f}_{V}^a\\
  \hat{\bf f}_{V}^a&=&-\imath\sigma_3{\bf f}_{X}^a\;.
                   \label{eq:bivectors}
\end{eqnarray}
We have
\begin{equation}
  \left(\hat{{\bf f}}_{X}^a,{\bf f}_{X}^b\right)=
  \left(\hat{{\bf f}}_{V}^a,{\bf f}_{V}^b\right)=M\delta_{ab}
\end{equation}
and
\begin{equation}
  \begin{array}{lll}
  \left(\hat{{\bf f}}_{E},{\bf f}_E\right)=
  \left(\hat{{\bf f}}_{\theta},{\bf f}_{\theta}\right)=\frac{\D M}{\D E}\\[4pt]
  \left(\hat{{\bf f}}_{\theta},{\bf f}_E\right)=  \left(\hat{{\bf f}}_{E},{\bf f}_\theta\right)=0\;,
  \end{array}
\end{equation}
with
\begin{equation}
M=\int u^2 \D^3\mbfx\;
\end{equation}
and all the other scalar products are vanishing.

In the presence of the quantum noise ${\bf s}$, we assume that all the soliton parameters
are time-dependent, and using~(\ref{eq:solitonansatz})  and we have after~(\ref{eq:A:nonlocalSDE})
\begin{equation}
  \label{eq:lowestorder}
  \begin{array}{lll}
  \fl  \imath\sigma_3\partial_t {\bm \psi}+\Delta{\bm \psi}-U\ast(\psi\psi^+){\bm \psi}={\bf s}\\
   ={\bf f}_\theta\left(-\dot\theta+t\dot E+\frac{1}{2}V^a\dot X^a-\frac{V^2}{4}\right)
    +{\bf f}_E\dot E+{\bf f}_{X}^a\left(\dot X^a- V^a\right)+{\bf f}_{V}^a\left(-\dot V^a\right)
    \end{array}
\end{equation}
where the dot indicates the time derivative. Equations~(\ref{eq:lowestorder}) are valid at the lowest
order of perturbation, higher orders can be determined by radiative corrections to the
soliton profile.
By scalar multiplying by $\hat{{\bf f}}_{X^a}$  and $\hat{{\bf f}}_{V^a}$, we obtain the stochastic equations for the position and the velocity of the soliton
  \begin{eqnarray}
    M \dot X^a=M V^a+\left(\hat{{\bf f}}_{X}^a, {\bf s}\right)\label{eq:ODEperturbationX}\\[4pt]
    M \dot V^a=-\left(\hat{{\bf f}}_{V}^a, {\bf s}\right)\;.
    \label{eq:ODEperturbationV}
  \end{eqnarray}
  Equations~(\ref{eq:ODEperturbationX}) and (\ref{eq:ODEperturbationV}) describe the dynamics of the soliton position and velocity with quantum noise.  Seemingly, we get equations for $\theta$ and $E$. At the lowest order in $t$ and $V^a=X^a=0$ at $t=0$, we
  have
  \begin{eqnarray}
    M' \dot E&=&\left(\hat{{\bf f}}_{E}, {\bf s}\right)\\
    M' \dot \theta&=&-\left(\hat{{\bf f}}_{\theta}^a, {\bf s}\right)\;.
    \label{eq:Etheta}
  \end{eqnarray}
  being
  \begin{equation}
    \label{}
M'=\frac{dM}{dE}
\end{equation}
\section{Quantum-induced parameter diffusion and random walk}
We have for the perturbation vector ${\bf s}$,
\begin{equation}
  \label{eq:singlenoise}
  {\bf s}= \sqrt{\frac{\imath}{n_0}}
  \left(\begin{array}{lll}
    \xi_U\\\imath \xi_U^+
  \end{array}\right)u {\bf e}\;.
\end{equation}
As detailed in \ref{sec:appA}, by using~(\ref{eq:singlenoise}), (\ref{eq:ODEperturbationX}), (\ref{eq:ODEperturbationV}) and~(\ref{eq:bivectors}), we obtain
\begin{eqnarray}
  M \dot X^a&=&M V^a+F^a_X({\bf X},t)\label{eq:7}\\
  M\dot V^a &=&F^a_V({\bf X},t)\;
                \label{eq:RW}
\end{eqnarray}
where $F^a_X$ and $F^a_V$, with $a=1,2,3$, are stochastic terms acting on the position $X^a$ and velocity $V^a$ of the soliton with mass $M$.
We have
\begin{eqnarray}
  F_a^X&=&\frac{1}{\sqrt{n_0}}\int \rho({\bf x}-{\bf X})(x^a-X^a) \xi_+({\bf x},t) \D^3 \mbfx \\
  F_a^V&=&\frac{1}{\sqrt{n_0}}\int \frac{\partial \rho}{\partial x^a}\left({\bf x}-{\bf X}\right) \xi_+({\bf x},t)\D^3\mbfx\;
\label{eq:12}\end{eqnarray}
 with ${\bf X}=(X^1,X^2,X^3)$ the soliton position, $\rho=u^2$, and $\xi_{+}(\mbfx,t)$ a real noise such that
 \begin{equation}
 \langle \xi_+({\bf x}',t') \xi_+({\bf x},t)\rangle=-U({\bf x}-{\bf x}')\delta(t-t')\;.
 \end{equation}
For the stochastic terms in Eqs.~(\ref{eq:RW}) we have
 \begin{eqnarray}
   \label{eq:FF}
   \langle F_X^a({\bf X},t) F_X^b({\bf X},t')\rangle &=&\frac{1}{n_0}Q_{X}^a\delta_{ab}\delta(t-t')\;,\\
   \langle F_V^a({\bf X},t) F_V^b({\bf X},t')\rangle &=&\frac{1}{n_0}Q_{V}^a\delta_{ab}\delta(t-t')\;,\\
   \langle F_X^a({\bf X},t) F_V^b({\bf X},t')\rangle &=&\frac{1}{n_0}Q_{XV}^a\delta_{ab}\delta(t-t')\;,
    \end{eqnarray}
    with the correlation coefficients
    \begin{eqnarray}
      \label{eq:QX}
      Q_{X}^a&=&-\int x_1^ax_2^a\rho(\mbfx_1)\rho(\mbfx_2)U(\mbfx_1-\mbfx_2)\D^3 \mbfx_1\D^3 \mbfx_2 \\
      Q_{V}^a&=&-\int\frac{\partial\rho(\mbfx_1)}{\partial x_1^a}\frac{\partial \rho(\mbfx_2)}{\partial x_2^a}U(\mbfx_1-\mbfx_2)\D^3 \mbfx_1 \D^3 \mbfx_2 \\
      \label{eq:QV}
      Q_{XV}^a&=&-\int x_1^a\rho(\mbfx_1)\frac{\partial \rho(\mbfx_2)}{\partial x_2^a}U(\mbfx_1-\mbfx_2)\D^3 \mbfx_1 \D^3 \mbfx_2 \;.
      \label{eq:QXV}
    \end{eqnarray}
    Equations~(\ref{eq:7}) and (\ref{eq:RW}) with $X^a(0)=V^a(0)=0$ give for the moments
    \begin{equation}
      \label{eq:V2}
\langle {\left[V^a(t)\right]}^2\rangle = D_V^a t\;,
    \end{equation}
    and
        \begin{equation}
      \label{eq:X2b}
\langle {\left[X^a(t)\right]}^2\rangle = D_X^a t+D_{XV}^a t^2+D_V^a\frac{t^3}{3}\;.
    \end{equation}
    The velocity and the position undergo a diffusive random walk with
    \begin{eqnarray}
      D_X^a&=&\frac{Q_X^a}{n_0 M^2}\\
      D_V^a&=&\frac{Q_V^a}{n_0 M^2}\\
      D_{XV}^a&=&\frac{Q_{XV}^a}{n_0 M^2} \;.
    \end{eqnarray}
    The diffusion in the position in~(\ref{eq:X2b}) arise from both the quantum
    noise and the diffusion of the velocity.
    At the lowest order  in $t$, we have
            \begin{equation}
      \label{eq:X2}
\langle {\left[X^a(t)\right]}^2\rangle = D_X^a t+D_{XV}^a t^2+D_V^a\frac{t^3}{3}\simeq D_X^a t\;.
    \end{equation}
    % The diffusion coefficients $D_{X}^a$, $D_{V}^a$, and $D_{XV}^a$   are computed numerically for a given interaction potential $U$.
    For the random walk of $\theta$ and $E$, we obtain after Eqs.~(\ref{eq:Etheta})
    \begin{equation}
      \label{}
      \begin{array}{lll}
\langle E(t)^2\rangle \simeq D_E t\\
\langle \theta(t)^2\rangle \simeq D_\theta t\\
      \end{array}
\label{eq:8}    \end{equation}
    with
    \begin{eqnarray}
      D_E&=&-\frac{4}{(M')^2 n_0}\int \rho(\mbfx_1) \rho(\mbfx_2) U(\mbfx_{1}-\mbfx_{2}) \D^3 \mbfx_1 \D^3 \mbfx_2\;, \\
      D_\theta&=&-\frac{1}{(M')^2 n_0}\int \rho'(\mbfx_1) \rho'(\mbfx_2) U(\mbfx_{1}-\mbfx_{2}) \D^3 \mbfx_1 \D^3 \mbfx_2\;,
      \end{eqnarray}
    being $\rho'(\mbfx)=\partial \rho(\mbfx)/\partial E$.

% \begin{eqnarray}
%                 \label{eq:RW}
%   M \dot X^a&=&M V^a+F^a_X({\bf X},t)\\
%   M\dot V^a &=&F^a_V({\bf X},t)\;
% \end{eqnarray}
% with ${\bf X}=(X^1,X^2,X^3)$ the soliton position, $F^a_X$ and $F^a_V$ stochastic terms, and $M=\int u^2\D^3 \mbfx$ the dimensionless soliton mass. We have in~(\ref{eq:RW})
% \begin{eqnarray}
%   F_X^a&=&\frac{1}{\sqrt{n_0}}\int \rho({\bf x}-{\bf X})(x^a-X^a)\xi_+({\bf x},t) \D^3 \mbfx
% \end{eqnarray}
% with  $ \xi_+=(\xi_U+\xi_U^+)/\sqrt{2}$, being $\langle \xi_+({\bf x}',t') \xi_+({\bf x},t)\rangle=U({\bf x}-{\bf x}')\delta(t-t')$.

%  The stochastic terms in Eqs.~(\ref{eq:RW}) are such that
%  \begin{equation}
%    \label{eq:FF}
%    \langle F_X^a({\bf X},t) F_X^b({\bf X},t')\rangle =\frac{1}{n_0}Q_{X}^a\delta_{ab}\delta(t-t')\;,\\
%     \end{equation}
%     with the correlation coefficient
%     %$\left[U_{12}=U(\mbfx_1-\mbfx_2)\right]$.
%     \begin{equation}
%       \label{eq:QQ}
%       Q_{X}^a=\int\int x_1^ax_2^a\rho(\mbfx_1)\rho(\mbfx_2)U(\mbfx_1-\mbfx_2)\D^3\mbfx_1\D^3\mbfx_2\;.
%     \end{equation}
% Results for $F_V^a$, $\theta$ and $E$ are given in the SM.

% \noindent Equation~(\ref{eq:RW}) gives, at the lowest order in $t$,
%   \begin{equation}
%       \label{eq:X2}
% \langle \left[X^a(t)\right]^2\rangle = D_X^a t;.
% \end{equation}
% with the diffusion coefficient  $D_X^a=Q_X^a/n_0 M^2$.
For the radially symmetric soliton, we show in figure~\ref{fig:solitonfluctuactions}b the diffusion coefficient $D_X^1=D_x^2=D_x^3=D_X$, as obtained by the numerical profile $u$ computed with the screened gravitational potential $U$. One finds that for a growing $\sigma$ the quantum diffusion is frustrated, as it happens in 1D~\cite{Folli10, Batz2011}. This can be deduced from (\ref{eq:QX}), indeed, as $\sigma\rightarrow\infty$, one has $U(\mbfx_{1}-\mbfx_{2})\simeq$~constant, and $Q_{X}^a\rightarrow 0$,
as for the soliton profile $\rho(\mbfx)=\rho(-\mbfx)$.

We compare (\ref{eq:X2}) with the full 3D+1 stochastic simulations and we find excellent agreement, as shown in figure~\ref{fig:solitonfluctuactions}c-e where we report the dynamics of solitary waves with $n_0 M\simeq 10^6$ atoms.

The diffusion constant $D=\hbar D_X/2 m$ in physical units reads
\begin{equation}
  D=\left(\frac{Q_{X}^{2}}{2 n_{0 }M^{2}}\right)\frac{\hbar}{m}=\left(\frac{Q_{X}^{2}}{2 M}\right)\frac{\hbar}{N_{T}m}\;.
\label{eq:11}\end{equation}
In Eq.(\ref{eq:11}) equation $N_{T}=n_{0}M$ is the total number of particles, and $N_{T}m=n_{0}M m$ the total mass of the condensate. $Q_{X}^{2}/2 M$ is a numerical constant that depends on the profile of the soliton and $\Lambda$.
We find that quantum fluctuations vanish when $N_{T}\rightarrow\infty$ or $\hbar\rightarrow  0$.
In the original units of Eq.(\ref{eq:nonlocalSDE}), the diffusion constant can be also cast as
\begin{equation}
D=\frac{G m^2}{\hbar}\int\int \frac{x_1 \rho(\mbfx_{1})}{\int \rho\D^3\mbfx} \frac{x_2 \rho(\mbfx_{2})}{\int \rho\D^3\mbfx} \frac{e^{-|\mbfx_1-\mbfx_2|/\Lambda}}{|\mbfx_1-\mbfx_2|}\D^3\mbfx_1\D^3\mbfx_2\;.\label{eq:D}
\end{equation}
We remark that Eq.~(\ref{eq:D}) is written in the original physical units of Eq.(\ref{eq:nonlocalSDE}), such that in Eq.~(\ref{eq:D}) $x$ is a length and the dimensions of $D$ are $m^{2}/s$ in the MKS system. Equation~(\ref{eq:D}) shows the interplay of quantum and gravitational effects through the ratio $G m^2/\hbar$ and returns $D$ in terms of the measurable density profile $\rho(\mbfx)$.
\section{Non-Gaussian statistics}
\begin{figure*}[hbt]
  \includegraphics[width=\textwidth]{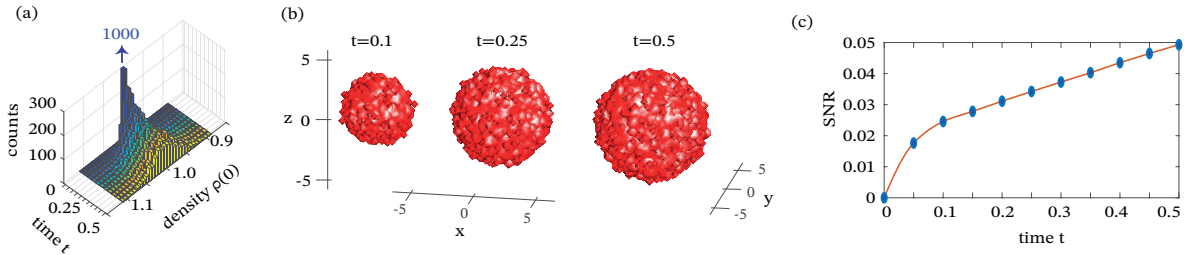}
  \caption{Time evolution and non-Gaussianity of the statistical distribution of 3D nonlocal solitons.
    (a) Histogram (after $1000$ runs) at different instants of the density $\rho$ computed at the classical soliton peak $\mbfx=0$;
    the vertical axis is truncated at $300$, the number of counts
    at $t=0$ and $\rho(0)=1.0$ is $1000$ as indicated by the arrow.
    (b) Temporal evolution of the spatial distribution of the non-Gaussianity parameter SNR of the
    density as 3D isosurfaces.  (c) Mean value of the SNR computed on the spatial profile versus time (parameters $\sigma=1, M=138$, $n_0=10^2$).
     \label{figure4}}
  \end{figure*}

In our stochastic simulations, the initial state is a coherent state, whose statistical properties change upon evolution. Here we follow~\cite{Howl2021} to determine if deviations from Gaussianity arise. We report in figure~\ref{figure4}a the evolution of the statistical distribution of the density $\rho(\mbfx=0)$ as computed by Eqs.~(\ref{eq:A:nonlocalSDE}) at the center of the classical solitonic core. The initial state is coherent, and the histogram is localized in the initial value of the peak. Upon evolution, the distribution spreads and manifestly displays a bell-shaped non-Gaussian profile. Similar behavior is also obtained for the quadratures of the field (not reported).

To quantify the deviation from Gaussianity, we consider the SNR introduced in~\cite{Howl2021}
  \begin{equation}
    \rm{SNR}=\frac{|\kappa_4|}{\sqrt{\rm{var}\,k_4}}
    \end{equation}
 here $\kappa_4$ is the fourth cumulant of the statistical distribution. $\rm{var}\,k_4$ is its uncertainty (see~\ref{sec:non-gauss-param}). For Gaussian statistics, all the cumulants higher than second order vanish, hence SNR measures deviation from non-Gaussianity including the uncertainty $\rm{var}\,k_4$ due to a finite number of samples. We compute SNR for the density and the field quadratures with similar results.

At variance with~\cite{Howl2021}, we account for the heterogeneous features of SNR, i.e., we measure SNR in different spatial locations. Figure~\ref{figure4}b shows the 3D isosurface of the SNR at different instants. The statistical distributions at different positions become non-Gaussian with time.  Figure~\ref{figure4}c shows the spatially averaged value of the SNR, which demonstrates that a self-trapped solitonic wave packet develops non-Gaussian statistics. Results in figure~\ref{figure4} refer to a representative case with $n_0 M\simeq 10^4$ atoms; we found these dynamics for different interaction lengths and particle numbers.

To understand the physical origin of the non-Gaussianity, we observe that - at the lowest order in $t$ - the soliton parameters $X^a$, $V^a$, $\theta$, and $E$, are the time-integral of white noise terms (i.e., Wiener processes). Thus they are the sum of many independent variables and hence obey Gaussian statistics. Non-Gaussianity arises from the fact that the soliton profile is a nonlinear function of these parameters, and any observable depends on the soliton profile. In general terms, the statistical distribution of a nonlinear function of a Gaussian
 variable is expected to be non-Gaussian. Thus, as far as the soliton is stable with respect to fluctuations,  non-Gaussianity arises. Nonlocal solitons are stable self-trapped nonlinear waves, and their robustness against quantum fluctuations induces non-Gaussianity.
\section{Summary}
In conclusion, we studied theoretically and by first-principle numerical simulations the 3D+1 dynamics of non-local self-gravitating boson fluids. The quantum noise induces diffusion in the self-localized wave-packet position determined by the degree of nonlocality and the particle number. The theoretical results agree with ab-initio 3D+1 simulations with no fitting parameters.

The quantum diffusion is due to the interplay of the quantum fluctuations and the long-range self-interaction. This interplay causes non-Gaussian statistics that spread in the solitonic core upon evolution. We remark that this is a universal phenomenon that is not dependent on the specific interaction potential $U$ but arises from the general stability properties of solitons.

Experimental investigations may involve long-range Bose-Einstein condensates (see, e.g.,\cite{Calvanese14} and references therein), and also nonlinear optical systems, where low-dimensional reductions of the Schr\"odinger-Newton equation have been considered~\cite{Rogers2016}.

The results open the way to using non-Gaussian multidimensional solitary waves as non-classical reservoirs for continuous-variable quantum information and as quantum simulators for quantum gravity models. Notably enough, the numerical simulations suggest that signatures of a quantized gravity may arise even without careful preparation of the initial state as Schr\"odinger cat (or squeezed state), but starting from a coherent solitonic state. Also, the results show the relevance of quantum fluctuations in cold dark-matter models, which can potentially impact the investigation of self-gravitating BEC and enable tests within astrophysical observations.

\ack
We acknowledge support from the H2020 PhoQus project (Grant no.820392).

\section*{Data availability statement}
All data that support the findings of this study are included within the article (and any supplementary files).

\appendix
\section{Stochastic equations for the soliton parameter}\label{sec:appA}
The perturbation vector is written as
\[
  {\bf s}=\frac{1}{\sqrt{n_{0}}}
\left(  \begin{array}{ccc}
    \sqrt{\imath}\xi_{U}\\\sqrt{-\imath}\xi_{U}^{+}
    \end{array}\right)u{\bf e}\;,
\]
where the two independent complex noises $\xi_{U}$ and $\xi_{U}^{+}$ are such that
\begin{equation}
\langle\xi_{U}(\mbfx,t)\xi_{U}(\mbfx',t')\rangle=U(\mbfx-\mbfx')\delta(t-t')
\label{eq:3}
\end{equation}
and
\begin{equation}
  \langle\xi^{+}_{U}(\mbfx,t)\xi^{+}_{U}(\mbfx',t')\rangle=U(\mbfx-\mbfx')\delta(t-t')\;,
 \label{eq:4} \end{equation}
being $U(\mbfx)=-\exp(-r/\sigma)/r<0$. We let
\begin{equation}
  \begin{array}{lll}
    \xi_{U}(\mbfx,t)&=&\imath C(\mbfx)\xi(\mbfx,t)\\
    \xi^{+}_{U}(\mbfx,t)&=&\imath C(\mbfx)\xi^{+}(\mbfx,t)\\
    \end{array}
  \end{equation}
  which satisfy (\ref{eq:3}) and (\ref{eq:4}),
  with $C(\mbfx)$ a real-valued function such that
  \begin{equation}
    C(\mbfx)\ast C(-\mbfx)=\int \D^{3}\mbfx' C(\mbfx-\mbfx')C(-\mbfx')=-U(\mbfx)>0\;,
  \end{equation}
  or, equivalently,
  \begin{equation}
   \int C(\mbfx-\mbfx')C(\mbfy-\mbfx')\D^{3}\mbfx'=-U(\mbfx-\mbfy)>0 \;.
    \end{equation}
Following Eq.(\ref{eq:ODEperturbationX}), we need the scalar product  $\left(\hat{{\bf f}}_{X}^a, {\bf s}\right)$ at $X^{a}=0$, that is
    \begin{eqnarray}
        \sqrt{n_{0}}\left(\hat{{\bf f}}_{X}^a, {\bf s}\right)&=2\Re\int \D^{3}\mbfx\left[-\imath\sigma_{3}
        \left(\hat{\bf f}_{V}^a\right)^{*}\right]\cdot {\bf s}\nonumber\\
      &=\Re\int\D^{3}\mbfx\, x^{a}u^{2}(\mbfx)\left(-\imath\sqrt{\imath}\xi_{U}+\imath\sqrt{-\imath}\xi_{U}^{+}\right)\nonumber\\
      &=\Re\int\D^{3}\mbfx\, x^{a}u^{2}C\ast(\sqrt{\imath}\xi-\sqrt{-\imath}\xi^{+})\nonumber\\
      &=\int\D^{3}\mbfx\, x^{a}u^{2}C\ast\frac{\xi-\xi^{+}}{\sqrt{2}}\\
      &=\int\D^{3}\mbfx\, x^{a}u^{2}C\ast\xi_{-}\;,
    \end{eqnarray}
    with $\xi_{-}\equiv(\xi-\xi^{+})/\sqrt{2}$ a real noise such that $\langle\xi_{-}(\mbfx,t)\xi_{-}(\mbfx',t')\rangle=\delta(\mbfx-\mbfx')\delta(t-t')$.
Seemingly, we have in Eq.(\ref{eq:ODEperturbationV})
     \begin{equation}
       \sqrt{n_{0}}\left(\hat{{\bf f}}_{V}^a, {\bf s}\right)=\int \D V \left(-\frac{\partial u^{2}}{\partial x^{a}}\right)C\ast\xi_{-}\;.
       \end{equation}
To solve the resulting Ito stochastic equations we define $\xi_{+}=C\ast\xi_{-}$ [see~(\ref{eq:12})], and
     \begin{equation}
\fl       F_{X}^{a}(t)=\int\D^{3}\mbfx\,\left(\hat{{\bf f}}_{X}^a, {\bf s}\right)=\frac{1}{\sqrt{n_{0}}}\int\D^{{3}}\mbfx x^{a}u^{2}C\ast\xi_{-}=\frac{1}{\sqrt{n_{0}}}\int\D^{3}\mbfx\,x^{a}u^{2}\xi_{+}\;,
\label{eq:13}     \end{equation}
     and
     \begin{equation}
    \fl   F_{V}^{a}(t)=-\int\D^{3}\mbfx\,\left(\hat{{\bf f}}_{V}^a, {\bf s}\right)=\frac{1}{\sqrt{n_{0}}}\int\D^{3}\mbfx\,\frac{\partial u^{2}}{\partial x^{a}}\,C\ast\xi_{-}=\frac{1}{\sqrt{n_{0}}}\int\D^{3}\mbfx \frac{\partial u^{2}}{\partial x^{a}}\xi_{+}\;.
  \label{eq:14}\end{equation}
  Eqs.~(\ref{eq:ODEperturbationX}) and (\ref{eq:ODEperturbationV}) read
  \begin{equation}
\eqalign{M \dot X^a=M V^a+F_{X}^{a}(t)\\
    M \dot V^a=F_{V}^{a}(t)\;.}
    \label{eq:ODEperturbationA}
  \end{equation}
  Eqs.(\ref{eq:ODEperturbationA}) are solved by quadratures as follows
  \begin{eqnarray}
    \fl M V^{a}(t)=\int_{0}^{t}F_{V}^{a}(s)\D s\label{eq:9}\\
\fl M X^{a}(t)=M\int_{0}^{t}V^{a}(s)\D s+\int_{0}^{t}F_{X}^{a}(s)\D s=\int_{0}^{t}\int_{0}^{s}F_{V}(u)\D u\D s+\int_{0}^{t}F_{X}^{a}(s)\D s\;.\label{eq:5}
  \end{eqnarray}
From (\ref{eq:9}) we have
\begin{equation}
  M^{2} \langle V^{a}(t)V^{b}(t')\rangle=\int_{0}^{t}\int_{0}^{t'}F_{V}^{a}(s)F_{V}^{b}(s')\D s \D s'\;.
\label{eq:10}\end{equation}
From (\ref{eq:5})
  \begin{eqnarray}
    M^{2}\langle X^{a}(t)X^{b}(t')\rangle &=\int_{0}^{t} \int_{0}^{s} \int_{0}^{t'}\int_{0}^{s'}\langle F_{V}^{a}(u)F_{V}^{b}(u')\rangle\D u \D u' \D s \D s'+\nonumber\\
                                          &+\int_{0}^{t} \int_{0}^{t'} \int_{0}^{s} \langle F_{V}^{a}(u)F_{X}^{b}(s')\rangle\D u \D s'\D s  +\nonumber\\
                                          &+\int_{0}^{t} \int_{0}^{t'} \int_{0}^{s'} \langle F_{V}^{a}(u')F_{X}^{b}(s)\rangle\D u' \D s'\D s+\nonumber\\
    &+\int_{0}^{t} \int_{0}^{t'} \langle F_{X}^{a}(s)F_{X}^{b}(s')\rangle \D s \D s' \label{eq:6}
  \end{eqnarray}

  We also have from Eqs.~(\ref{eq:13}) and~(\ref{eq:14}) the following
  \begin{eqnarray}
    \langle F_{V}^{a}(t)F_{V}^{b}(t') \rangle=Q_{V}\delta_{ab}\delta(t-t)\\
    \langle F_{X}^{a}(t)F_{V}^{b}(t') \rangle=Q_{XV}\delta_{ab}\delta(t-t)\\
    \langle F_{X}^{a}(t)F_{X}^{b}(t') \rangle=Q_{X}\delta_{ab}\delta(t-t)\\
\label{eq:15}\end{eqnarray}
where we accounted for the fact that $u^{2}(\mbfx)=u^{2}(-\mbfx)$, and (\ref{eq:QX}),(\ref{eq:QV}) and (\ref{eq:QXV}) hold.
By using~(\ref{eq:15}) in (\ref{eq:10}) and (\ref{eq:6}) and letting $t=t'$ we have Eq.~(\ref{eq:V2}) and Eq.~(\ref{eq:X2}).
Similar arguments lead to (\ref{eq:8}).
\section{Non-Gaussianity parameter}\label{sec:non-gauss-param}
    The fourth order cumulant $\kappa_{4}$ is computed by using the value of the density $\rho(\mbfx,t)$, or of the field quadratures. Denoting as $q$ a value of a single run,
    we first determine the non-central moments ($m=0,1,2,\ldots$)
    \begin{equation}
\mu'_m=\langle q^m\rangle\,.
    \end{equation}
    Then we compute the first $8$ cumulants $\kappa_n$ with $\kappa_1=\mu'_1$, and ($n>1$)
    \begin{equation}
\kappa_n=\mu_n'-\sum_{m=1}^{n-1}\kappa_{n-m}\mu_m'\;.
    \end{equation}
    For the $k-$statistics, we have $\langle k_4\rangle=\kappa_4$, and, letting $\mathcal{M}$ the number of runs,
    \begin{equation}
      \begin{array}{lll}
        \rm{var}(k_4)&=\langle (k_4-\kappa_4)^2\rangle=
        +\frac{\kappa_8}{\mathcal{M}}+16\frac{\kappa_2\kappa_6}{\mathcal{M}-1}
        +48\frac{\kappa_3 \kappa_5}{\mathcal{M}-1}+34\frac{\kappa_4^2 }{\mathcal{M}-1}+\\
        &72\frac{\mathcal{M}\kappa_2^2 \kappa_4}{(\mathcal{M}-1)(\mathcal{M}-2)}  +144\frac{\mathcal{M}\kappa_2 \kappa_3^2}{(\mathcal{M}-1)(\mathcal{M}-2)}
    +24\frac{\mathcal{M}(\mathcal{M}+1)\kappa_2^2}{(\mathcal{M}-1)(\mathcal{M}-2)(\mathcal{M}-3)}\;.
        \end{array}
    \end{equation}
\section*{References}
%%%%%%%%%%%%
%\bibliographystyle{unsrt}
\bibliographystyle{iopart-num}
%\bibliography{../../../../bibtex/GIGAbib}
\providecommand{\newblock}{}

%%%%%%%%%%%%%%%%%%%%%%%%%%%%%%%%%%%%%
\end{document}